\documentclass[9pt]{article}

\usepackage{setspace}
\usepackage{blindtext}
\usepackage[a4paper, total={7in, 8in}]{geometry}

\newcommand{\Z}{\mathbb{Z}}

\makeatletter
\newcommand*{\rom}[1]{\expandafter\@slowromancap\romannumeral #1@}
\makeatother

\usepackage{amsmath} 
\usepackage{amsfonts}
\usepackage{xcolor}
\usepackage{graphicx}

\date{}
\begin{document}

\title{Construction of Multiple Constrained DNA Codes}
\maketitle

\vskip6pt
\begin{center} 
{{
Siddhartha Siddhiprada Bhoi\footnote{Department of Mathematics \& Computing, Indian Institute of Technology (ISM), Dhanbad, India. {\tt siddhartha2006rta@gmail.com}},
P. Udaya\footnote{School of Computing and Information Systems, The University of Melbourne, Parkville, Australia. {\tt udaya@unimelb.edu.au}},
Abhay Kumar Singh\footnote{Department of Mathematics \& Computing, Indian Institute of Technology (ISM), Dhanbad, India. {\tt abhay@iitism.ac.in}}, 

}}
\end{center} 
\begin{abstract}
DNA sequences are prone to creating secondary structures
by folding back on themselves by non-specific hybridization
among its nucleotides. The formation of secondary structures
makes the sequences chemically inactive towards synthesis and
sequencing processes. In this letter, our goal is to tackle the
problems due to the creation of secondary structures in DNA sequences along with constraints such as not having a large homopolymer run length. In this paper,
we have presented families of DNA codes with secondary structures of stem length at most two and homopolymer run length at most four. By mapping the
error correcting codes over $\Z_{11}$ to DNA nucleotides, we obtained
DNA codes with rates $0.5765$ times the rate of corresponding code over $\Z_{11}$, which include some new secondary structure free and better-performing codes for DNA based data storage and DNA computing purposes.
\end{abstract}

\section{Introduction}
%
%
%
%
Deoxyribose Nucleic Acid(DNA) is nature's way of storing genetic data and has been the central concept in DNA computing\cite{a9}. DNA based data storage is important due to its high storage density, capacity, and longevity \cite{a1}, \cite{a43}. These advantages motivated researchers to explore the development of the subject. \\
Let $S_{DNA} = \{A,T,G,C\}$, a set of four DNA nucleotides named Adenine (A), Guanine (G), Cytosine(C), and Thymine (T). DNA is made from DNA strands, which are sequences over $S_{DNA}$ and for any positive integer $n$, a DNA code $S^n_{DNA}$ is a collection of DNA sequences of length $n$.
For each $X = x_1 \cdots x_n \in S^n_{DNA}$, then the reverse of the DNA sequence denoted by $X^r$ is $x_n x_{n-1} \cdots x_1$, the complement of the DNA sequence denoted by $X^c$ is $x_1^c x_2^c \cdots x_n^c$, and the reverse complement of the DNA sequence denoted by $X^{rc}$ is $x_n^c x_{n-1}^c \cdots x_1^c$. The complements of the DNA nucleotides are $T^c = A$, $A^c =T$, $G^c= C$, and $C^c=G$.  DNA sequences are synthesized physically and read by the methods of DNA synthesis and DNA sequencing, respectively \cite{a21}. During the synthesis and sequencing, various types of errors can occur, which bring in the importance of the study of the minimum distance of DNA code, the property that enables the suppression of those errors. \par

The purpose of DNA sequencing is to read the DNA content,
i.e., to determine the exact DNA nucleotides along with their order. This is accomplished using the specific hybridization between the DNA sequence and its complement \cite{a3}. Hybridization is the process of complementary base pairs binding to form a double helix. The main source of errors at the time of the sequencing process is due to non-specific hybridization. If in a DNA code some of the DNA sequences are similar enough among themselves or their reverse or reverse-complement versions,
then non-specific hybridization will take place. This situation motivates authors [\cite{a3}, \cite{a4}, \cite{a5}, \cite{a50}, \cite{ab52}] to design DNA codes whose sequences are sufficiently different among themselves as well as from their reverse and reverse complement versions.
The reverse-complement constraint is the constraint in which DNA sequences and their reverse complement versions are far apart by at least the minimum Hamming distance of the code.\par

GC content is the percentage of nitrogenous bases in a DNA molecule that are either guanine ($G$) or cytosine ($C$). The average GC-content in human genes ranges from $35-60$ percent across $100$-Kb fragments \cite{ab3}. The variation in GC content can lead to variation in melting temperature and stability of the DNA strand. DNA strands with $50$ percent GC content have the highest stability, thus GC content is kept to nearly $50$ percent for the DNA based data storage.  \par

At the time of synthesis and sequencing of DNA sequences with a high repetition of nucleotides, insertion, deletion, and substitution errors can occur \cite{a43}. This high consecutive repetition of nucleotides in DNA sequences is called a homopolymer run. A DNA sequence is free from a homopolymer of run length $l$ if all nucleotides in any subsequence of length $l$ are not the same. For example, the DNA sequence $ACGCCCCCGTG$ has a homopolymer of run length $6$. In \cite{ab52}, the authors have discussed enumerating methods and bounds over DNA sequences with GC constraints, homopolymer run length at most 3,  and specific Hamming distances.  \par
In a DNA strand, the nucleic acids can have primary, secondary and tertiary structures. The primary structure consists of a linear sequence of nucleotides, the secondary structure is due to the base-pairing interaction of nucleotides within an active single-stranded DNA sequence \cite{a9}, and tertiary structure is due to the position of atoms in 3D space and its steric and geometric constraints. A DNA strand can fold back upon itself and form a loop-like structure with itself. Because of this structure, DNA becomes slow to react to chemical reagents in the sequencing process\cite{a34}. Therefore, to read such DNA sequences with a secondary structure, the DNA needs to be unfolded, which can require extra resources and energy, leading to storage inefficiency. The authors \cite{a9} reports new DNA codes avoiding secondary structure. For some particular cases, some constraints such as reverse and reverse-complement constraints and fixed GC-content constraints are also discussed in   \cite{a9}, \cite{a4}.\par
Benerjee et.al.\cite{a0} took an interesting approach to the code construction by considering a certain set of DNA sequences of length two as the base alphabet and mapping it to the field $\Z_5$. This method enabled them to define DNA codes as an image of codes over $\mathrm{\Z_5}$. It is a natural question to find if other algebraic structures can provide a map to DNA codes with better parameters for DNA based data storage such that the resulting DNA sequences avoid secondary structure and satisfy the other necessary constraints.  

After providing the required notations and preliminaries in Section II, the rest of the letter is organised as follows.  Homopolymers and Secondary structure avoiding maps, and their properties are discussed in the next section. Further, in Section IV, families of DNA codes are constructed, and other families of codes like Hamming codes and Reed Solomon codes are studied. In Section V, a comparative analysis is established, and a comparative table of code rates and code attributes is provided.
\section{PRELIMINARIES AND NOTATION}
Let $F_q$ denote an alphabet of size $q$ symbols. For any positive integer $n$, a code $C$ over the alphabet $F_q$ is a subset $C \subset F_q^n$ of size $M$ and minimum distance $d$. The Hamming distance for any two sequences of equal length $x$ and $y$ is denoted by $d_H(x,y)$ and defined  as the number of positions at which the corresponding symbols are different. The Hamming distance for a code $C$ is defined as $d_H(C) = \mathrm{min} \{ d_H(x,y): x \neq y $ $  \forall $ $ x,y \in C \}$. A DNA code $C_{DNA}$ is any code defined over the alphabet $S_{DNA}$. For any DNA code $C_\mathrm{DNA}$ with minimum
Hamming distance $d_H$, the reverse minimum distance $d^r_H= \mathrm{min}\{d_H(x,y^r):x \neq y^r$ and $x, y^r \in C_\mathrm{DNA}\}$.  Let $X=x_1 \cdots x_n \in S_\mathrm{DNA}^n$ be a DNA sequence of length $n$, then for any integer  $i,j$ such that $1 \leq i \leq j\leq n$ the continuous sequence $x_i,x_{i+1} \cdots x_j$ is called a subsequence of $X$.\par
We have defined free energy and interaction energy and also provided the relation between them. We have also discussed how the free energy affect the secondary structure. A chemically active DNA sequence $ x_1x_2 \cdots x_n$,  to gain stability forms secondary structure and it releases energy known as \textit{free energy} \cite{a34}, denoted by $E_{1,n}$. The free energy of a DNA sequence can be used to predict the secondary structure of the DNA sequence. The free energy depends on the energy released by pairing $(x_i,x_j)$, where $1 \leq i < j \leq n$. This released energy is called \textit{interaction energy}. The Nussinov-Jackson folding algorithm (NJ algorithm) is used to approximately predict the secondary structure in a given DNA sequence \cite{a12}. In the NJ algorithm, it is assumed that the interaction energy $\mu (x_i,x_j)$ is independent of all other nucleotide pairs. They only depend on the chosen nucleotide base pair $(x_i,x_j)$. Commonly used values of $\mu (x_i,x_j)$ are
\[\mu (x_i,x_j) = 
\left\{
\begin{array}{ll}
      -5 & if (x_i,x_j)\in \{(C,G),(G,C)\}, \\
      -4 & if (x_i,x_j)\in \{(T,A),(A,T)\}, \\
      -1 & if (x_i,x_j)\in \{(T,G),(G,T)\}, \\
       \ \ 0 & otherwise \\
\end{array} 
\right. \]
As NJ algorithm consider interaction energy ,$\mu (x_i,x_j)$, to be independent of other nucleotides except $x_i$ and $x_j$, we can calculate minimum free energy of a DNA sequence through a recursion. 
The minimum free energy for a subsequence $x_ix_{i+1}\cdots x_j$ of a DNA sequence $x_1x_2 \cdots x_n$ is given as $E_{i,j} = \mathrm{min}\{E_{i+1,j-1}+\mu (x_i,x_j), E_{i,k-1}+E_{k,j} : k = 1,2,\cdots,j\}$ with initial conditions $E_{l,l} = E_{l-1,l} = 0$ for $l=1,2,\cdots,n$ \cite{a9}. A larger negative value for free energy means a higher probability of secondary structure  in the DNA sequence. The difference in the values of the free energies is due to the relative strength of the bonds between DNA nucleotides, i.e., Guanine(G) and Cytosine(C) have a triple bond between them whereas Adenine(A) and Thymine(T) have a double bond between them. So, the relative strength of the bond between Guanine(G) and Cytosine(C) is higher than that of Adenine(A) and Thymine(T).
  For a DNA sequence $x = x_1 x_2\cdots x_n$, we define secondary-complement sequence as $x^{sc}  = x_n^{sc}  x_{n-1}^{sc}\cdots x_1^{sc}$  with $A^{sc}$ is $T$, $C^{sc}$ is $G$, $T^{sc}$ can either $A$ or $G$ and $G^{sc}$ can either be $C$ or $T$. 
Note that the reverse complements of a DNA sequence are secondary complements, but the converse is not true in general. A DNA subsequence can have multiple secondary complements i.e., $(ATCA)^{sc}=TGAT$ and $(ATCA)^{sc}=TGGT$. If we take two disjoint subsequences of a given DNA code such that one is a secondary complement of the other, then the subsequences are called disjoint secondary-complement subsequences of the DNA sequence. The goal of the paper is to find such DNA sequences which are stable without secondary structure. \par
\begin{figure}
\includegraphics[scale=1]{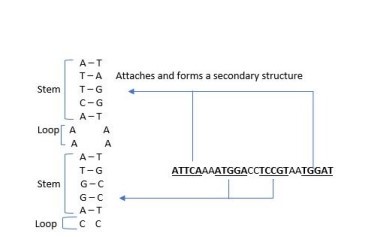}
\end{figure}
\par
Let the nucleotides in a DNA sequence $x = x_1x_2 \cdots x_n \in S_\mathrm{DNA}^n$ be indexed from 1 to n and for any $1\leq i,j\leq n$, $(x_i,x_j)$ denote the pairs contributing to the secondary structure. These are the pairs of nucleotides that have negative interaction energy as per NJ algorithm. For any $1 \leq i_1, j_1,i_2,j_2 \leq n$, the pairs of nucleotide  $(x_{i_1}, x_{j_1})$  and $(x_{i_2},x_{j_2})$ are said to be a consecutive set of pairs of nucleotide if both $i_1$ and $i_2$ are consecutive and $j_1$ and $j_2$ are consecutive. We define the stem length of a DNA sequence as the number of a consecutive set of pairs of nucleotides contributing to secondary structure and stem is defined as the consecutive set of pairs contributing to secondary structure. We consider these since many pairs are not possible due to chemical and stereo-chemical constraints \cite{a10}. \par
For example, in a DNA sequence $ATTCAAAATGGATCCGTAATGGAT$, $\{(x_i,x_{25-i})$, $(x_j,x_{5+j})  | i= 1,2,3,4,5; j=8,9,10,11,12 )\}$ are two disjointed secondary structures of length $5$ as shown in Fig 1. As $(ATTCA)^{sc}=TGGAT$ and $(ATGGA)^{sc}=TCCGT$.\\ 
 \textbf{Lemma 1} \textit{ For a given positive integer $l$, if the DNA sequence is free from a disjoint secondary complement of length $l+1$, then there cannot exist any secondary structure of stem length more than $l$.}\\ 
\textsc{Proof.} It is observed from the free energy of a DNA sequence that if there exist two disjoint secondary-complement sub-sequences $\alpha$ and $\beta$ of length $l+1$ in the DNA sequence such that $\alpha$ = $\beta^{sc}$ then there exists a secondary structure of stem length $l+1$. It is also observed that if the DNA sequence is free from secondary-complement sub-sequence of length $l$, it is also free from secondary-complement sub-sequence of length $l+k$ where $k$ is a non-negative integer. The lemma follows from the contrapositive of the above observations.   \hfill $\square$\\

For example, consider the DNA sequence $X= ATTCAAAATGGATCCGTAATGGAT$ as shown in Fig 1. The sequence $X$ have two disjointed secondary structures of length $5$ and hence it is free from a disjoint secondary complement of length $6$. So according to Lemma 1, $X$ can not have a secondary structure of stem length more than $5$.\\
\section{DNA Alphabets and Their Properties}
\noindent
The presence of secondary structure in a DNA sequence depends upon the value of free energy of the DNA sequence and the value of free energy depends upon the relative bond strength and bonding between some of the  DNA subsequences. By avoiding such subsequences, we can control the free energy of DNA sequence and hence have control upon the presence of secondary structure in the DNA sequence.\par
The structure of the base alphabet is extremely important to control
the free energy, which depends on the interaction between the nucleotides. If the base alphabet is $S_\mathrm{DNA}$, then in the pseudo-random sequence method, it is impossible to control the secondary structures as interactions are ensured between every possible combination due to the nature of the sequence generation mechanism. In \cite{a0}, the authors choose a base alphabet as $\{AA,AC,CA,CC,TC\}$, a subset of length two nucleotides. In this letter, we consider the base alphabet of length $3$. 
We are interested in a set S to be the set of possible base
alphabets of length $3$ nucleotides which can avoid secondary structure when DNA sequences are constructed as concatenations of the elements in the base alphabet. The set $ S= \{CCC, CCA, CAC, CAA, ACC, ACA, AAC, AAA, TCC,$ $ CTC, TCA\} $ with 11 elements is considered for this letter. For any DNA code $C_\mathrm{DNA} \subseteq  S^n$ of length $3n$ constructed using a bijective map $\Z_{11}$ to $S$.  We use $\Z_{11}=(0, 1, 2, 3, 4, 5, 6, 7, 8, 9, \text{\rom{10}})$ in our code construction, where $\text{\rom{10}} = 10 \mathrm{\mod} 11$.\\ 
The bound on free energy on DNA codes over $S^n$ is obtained through mathematical induction on the parameter $n$ for the free energy presented below in proposition 1.
\\
\textbf{Proposition 1.} \textit{For any DNA sequence in $S^n$, $E_{1,3n} \geq -4n$}.\\ 
Proof of the above Proposition is provided in the appendices.\\

It should be noted that for any arbitrary DNA sequence of length $3n$, we have $E_{1,3n} \geq -5\left \lfloor 3n/2 \right \rfloor$. From Proposition 1, it is clear that the free energy for any DNA sequences over $S$ is restricted thus avoiding secondary structures as explained in \cite{a0}.\\ 
\textbf{Lemma 2} \textit{For positive integers $n$, $m$, and $l$, such that $n\geq 6$ and $2\leq l\leq \left \lfloor m/2 \right \rfloor$. The set $U_n = \{a_1a_2\cdots a_n:a_i\in U \subset S^m_\mathrm{DNA}, i = 1,2,\cdots,n\}$. If each DNA sequence in $U_{6}$ is free from l length secondary-complement sub-sequences, then each DNA sequence in $U_n$ is also free from disjoint secondary-complement sub-sequences of length more than l.  Any 3n-length DNA sequence in $S^n$ is free from a secondary structure of stem length more than one.}\\ 
\textsc{Proof.} By the way of construction of $U_{6}$, DNA sequences of $U_{6}$ ensure that concatenation of any six DNA sequences from $U$ is free from $l$ length secondary-complement sub-sequences. So, the same stands true for n DNA sequences from $U$. For m = 3; l = 2, the results follow. \hfill$\square$\\
 Note that Pairing between G and C, and between G and T is not possible, thus preventing many secondary structures. Again, if any secondary structure exists, then it cannot have a stem length of more than one, and thus less energy is required to break it.
From Lemma 2 and Remark, it is clear that any DNA sequences in $S^n$ are free from a secondary structure of stem length of more than one.
\subsection{Secondary Structure Avoiding Map}
Define a bijective map $\phi: \Z_{11} \to S$ such that:
\[  {\begin{tabular}{cccccc}
x & $\boldmath{\phi(x)}$ & x & $\boldmath{\phi(x)}$ & x & $\boldmath{\phi(x)}$ \\
0     & CCC & 4 & ACC & 8     & TCC\\
1     & CCA & 5 & ACA & 9     & CTC \\
2     & CAC & 6 & AAC &  \rom{10}    & TCA \\
3     & CAA & 7 & AAA &     &  \\
  \end{tabular} } \]
$ $
where $x \in \Z_{11}$ and $\phi(x) \in S$. For any sequence $x = x_1x_2\cdots x_n \in \Z_{11}^n$, we have $\phi(x) = \phi(x_1)\phi(x_2)\cdots \phi(x_n)$. For any $x$ and $y$ $\in \Z_{11}$, we define distance function as $d:\Z_{11} \times \Z_{11} \to R$ such that $d(x,y) = d_H(\phi(x),\phi(y))$. For any $x$ and $y$ $\in \Z_{11}^n$, we define $d(x,y) = \sum_{i=1}^n d(x_i,y_i)$. For any code $C$ over $\Z_{11}$, $d = \mathrm{min}\{d(x,y): x \neq y \ and\  x,y \in C\}$.\\ 
\textbf{Proposition 2.}  \textit{$(\Z_{11}^n,d)$ and $(S^n,d_H)$ are isometric.}\\ 
\textsc{Proof.} The function $\phi :(\mathbb{Z}_{11}^n,d) \rightarrow (S^n,d_H)$  is an isometric function as it preserves the distance functions $d$ and $d_H$  by its definition $d(x,y)=d_H(\phi(x),\phi(y))$ for all $x,y \in \mathbb{Z}_{11}^n$.\hfill$\square$  \\
\textbf{Theorem 1.} \textit{For every code $C$ with parameters $(n,M,d)$ over $\Z_{11}$, there exists a DNA code $\phi (C)$ with parameters $(3n,M,d_{H})$ such that each DNA sequence in $\phi (C)$ is free from secondary structure of stem length more than one and the minimum Hamming distance $d_{H}=d$.}\\ 
\textsc{Proof.} From Proposition 2, $\phi$ is an isometric from $(\Z_{11}^n,d)$ to $(S^n,d_H)$. Thus, they have the same size and Hamming distance. The proof follows from Lemma 2.\hfill$\square$\\ 
\textbf{Lemma 3.} \textit{$d_H(x,y^c)\geq n$ for any $x,y \in S^n$}.\\ 
\textsc{Proof.} It is observed that for any $x,y\in S$, $d_H(x,y^c)\geq 1$. This implies $d_H(x,y^c) = \sum_{i=1}^{n}d_H(x_i,y^c_i)\geq n$.\hfill$\square$\\ 
\textbf{Lemma 4.} \textit{For a $(3n,M,d_H)$  DNA code $C$ over $S$, if $d_H \leq n$ then $C \cup C^c$ is a $(3n,2M,d_H)$ DNA code where $C^c = \{x^c:x\in C\}$}. \\
\textsc{Proof.} From the definition of complement of DNA sequence the results on the size and length follows. For any DNA code $C_{DNA}$ over $S$,let $x,y \in C_{DNA} $ be two DNA sequences. Then $d_H(x^c,y^c)=d_H(x,y) \geq d_H$ and from Lemma 3 $d_H(x^c,y^c)=d_H(x,y) \geq n$. So, for $C_{DNA} \cup C_{DNA}^c$, the minimum Hamming distance is $\mathrm{min} {d_H,n}=d_H$. Hence,the result on distance follow. \hfill$\square$\\
\textbf{Lemma 5.} \textit{For any $(n, M, d)$ code $C\subset \Z_{11}^n$, $\phi(C)$ exists and if $d\leq n$ then it satisfies the reverse complement constraint.}\\ 
\textsc{Proof.} If $x,y \in S^n$, $d_H(x,y^{rc})\geq n \geq d$ and thus the proof follows from Proposition 2.\hfill$\square$ 
\subsection{Homopolymers Deletion Map}
Define a non-surjective map $f:S^n\to S_\mathrm{DNA}^{3n}$, such that for a sequence $y \in S^n$, there exists a subsequence with consecutive repeated nucleotides of length more than four, then we flip the nucleotide present in doubly even positions of $y$ to $T$.\par For $f(y) = y_1y_2y_3\cdots y_{3n}$ if $y_{i+1}\cdots y_{i+l}$ are of same nucleotide with $l\geq 4$, then flip $y_{i+j}$ to $T$ if $j \mod 4 =0 $. For example, $f(ACCACCCCCCCAAAAAAA) = ACCACCCTCCCTAAATAAA$. It does not contain any homopolymers of run length more than four.\\ 
\textbf{Lemma 6.} \textit{For any DNA sequence $x \in S^n$, $f(x)$ cannot have a secondary structure with a stem length more than two.}\\ 
\textsc{Proof.} For any DNA sequence $x \in S^n$, we observe that in $f(x)$ the flipped nucleotides varies at least by four. Thus, from Lemma 2, the stem length of a secondary structure cannot be more than two.\hfill$\square$ 
  \\ 
\textbf{Theorem 2.} \textit{For every code $C$ over $\Z_{11}$ with parameters $(n,M,d)$, there exists a reverse-complement DNA code $f(\phi(C))$ with parameters $(3n,M,d_H)$ such that each DNA sequence in $f(\phi(C))$ is free  from homopolymers of run length more than four and from secondary structure of stem length more than two, where $\left \lceil \frac{3d}{4} \right \rceil \leq d_H \leq \left \lceil \frac{3n+3d}{4}  \right \rceil$}.
\\ 
\textsc{Proof.} If there exist $m$ consecutive nucleotides, then at most $\left \lfloor m/4 \right \rfloor$ nucleotides are flipped to $T$. Thus for any $x,y\in S^n$ $d_H(x,y) - \left \lfloor m/4 \right \rfloor \leq d_H(f(x),f(y)) \leq d_H(x,y) + \left \lfloor m/4 \right \rfloor$. The upper and lower bounds follow from the extreme cases, i.e.,  $m = 3n-d_H$ and $m = d_H$.\hfill$\square$
\section{Families of new constrained DNA codes}
\subsection*{Family 1:}
For $q=11$, consider the family of codes $C_k$ of length $4^{k-1}$ and dimension ${k}$ with $2 \leq k \leq 5$ with generator matrix $G_k$. The matrix
$$G_2 = \left(\begin{matrix}
1 & 1 & 1 & 1 \\ 
1 & 5 & 9 & \text{\rom{10}}
\end{matrix}\right)
$$
and
$$G_{i+1} = \left(\begin{matrix}
1_{4^{i-1}} & 2_{4^{i-1}} & 3_{4^{i-1}} & 4_{4^{i-1}} \\ 
G_i & G_i & G_i & G_i
\end{matrix}\right), \text{ for}\ \ i = 2,3,4.
$$
where $i_j$ is all the $i$ vectors of length $j$ for $i = 2,3,4,5$.
The code meets the Griesmer bound: $n = \sum_{i=0}^{k-1}\left \lceil \frac{d}{q^i} \right \rceil$. 

\textbf{Theorem 3.} \textit{There exists a $(2^{2k-1}, 5^k,  3\cdot4^{k-2})$ DNA code $f(\phi(C_k))$ with $d_H^r=2^{2k-3}$}.\\ 
\textsc{Proof.}  The distance $d(0_4,4190)=3$, where $4190 = 2(1_4+159$\rom{10}$)$ is in the linear span of $G_2$. Thus $d(0_{4^{k-1}},(4190)^{k-1}) = 3\cdot 4^{k-2}$ for any positive integer k. From symmetry, the minimum distance for the code $C_k$ is $d(0_{4^{k-1}},(4190)^{k-1})=3\cdot 4^{k-2}$. The size and length follow from Theorem 1.\hfill$\square$
\subsection*{Hamming Codes:}Consider a q-ary Hamming code. For an integer r$(\geq 2)$, the Hamming code is a linear code with parameters $\left [\frac{q^{r}-1}{q-1},\frac{q^{r}-1}{q-1}-r,3\right ]$ over $\mathbb{F}_q$. For $q = 11$ we have the following results:\\  
\textbf{Theorem 4.} \textit{There exists a $\left (\frac{3(11^{r}-1)}{10},11^{\frac{11^{r}-1-10r}{10}},3\right )$ DNA code $\phi(H_r)$ for Hamming code $H_r$ over $\mathbb{Z}_{11}$}.\\
\textsc{Proof.} The Hamming code over $\mathbb{Z}_{11}$ of degree $r(\geq 2)$ is a $\left [\frac{11^{r}-1}{10},\frac{11^{r}-1}{10}-r,3\right ]$ linear single error correcting code. The size and length of the code follow from the parameters of $H_r$ and from Theorem 1. For some $ x= 0_{\frac{11^{r}-1}{10}}$ and $y=0_{\frac{11^{r}-31}{10}}191$ are the codewords in $H_r$ such that $d(x,y)=3$, where $0_i$ represents all zero vector of length i. As $3=d_H^* \leq d$, it follows the result.\hfill$\square$
  \\ 
\textbf{Lemma 8.} \textit{For the code $H_r$ over $\mathbb{Z}_{11}$, there exists a $\left (\frac{3(11^{r}-1)}{10},11^{\frac{11^{r}-1-10r}{10}},d_H^*\right )$ DNA code $f(\phi(H_r))$ for $r=2$, $d_H^* = 3$}. \\
\textsc{Proof.} The proof follows from Theorem 2 and Theorem 4.\hfill$\square$
\subsection*{Reed Solomon Codes:}
A q-ary Reed Solomon code is a q-ary Bose, Chaudhari, and   Hocquenghem (BCH) code of length $q-1$ generated by 
$$g(x) = (x-\alpha^{a+1})(x-\alpha^{a+2})\cdots (x-\alpha^{a+\delta -1}),$$
where $a \geq 0 $, $2 \leq \delta \leq q-1$ and $\alpha$ is a primitive element of $q$-ary field. Reed-Solomon codes are  Maximum Distance Separable (MDS) codes with parameters $[q-1, q-\delta, \delta]$ over $F_q$, for any $2 \leq \delta \leq q-1$. As Reed Solomon Codes have high reliability and increase in code rate with increase in alphabet size, they are used widely for Data Storage and long distance communication purposes. So for $\Z_{11}$,  RS codes have the following properties:\\ 
\textbf{Theorem 5.} \textit{There exists a $\left (30,11-\delta,\delta \right )$ DNA code $\phi(RS)$ for Reed Solomon Code $RS$ over $\mathbb{Z}_{11}$}. \\
\textsc{Proof.} The Reed Solomon code over $\mathbb{Z}_{11}$ is a $\left (10,11-\delta,\delta \right)$ linear MDS code. The size and length of the code follow from the parameters of $H_r$ and Theorem 1.\hfill$\square$
\\ 
\textbf{Lemma 9.} \textit{For the family of Reed Solomon code over $\mathbb{Z}_{11}$, there exists a single error correcting code of parameter $\left (30,8,3\right ).$} \\
\textsc{Proof.} The size and length of code follow from Theorem 2 and Theorem 5 for $\delta= 3$. $0020010001$ is a $RS(11)$ codeword and $d(0_{10},0020010001) = 3$. \hfill$\square$ \\
\section{Discussion of the results}
It is observed that all DNA codes constructed in Section IV are reverse-complement DNA codes which can also be concluded from Lemma 6 and Theorem 2. In Theorem 3 and Theorem 4, the DNA codes are free from the secondary structure with stem length starting from one. Similarly, the DNA codes in Lemma 8 and Lemma 9 are free from homopolymers of run lengths more than four and secondary structures of stem length more than two. Comparisons of properties of single error-correcting DNA codes are made in Table 1 against the properties, $P_1$ to $P_3$, are respectively the secondary structure avoiding property, the homopolymers run length, and reverse complement property, respectively. The DNA codes constructed from  Reed-Solomon codes over $\Z_{11}$ appear to show a comparatively better code rate than other codes in the literature. While comparing with the codes presented in \cite{a0}, for some lengths of DNA sequences, our codes show a higher code rate. This also opens the problem of finding appropriate algebraic structures to construct DNA codes that can provide optimum code rates while obeying all the constraints of DNA codes.\par
In this letter, we have found DNA codes of a certain length providing a higher code rate than the DNA codes of similar error correcting capabilities provided in the literature. For example, a single error Reed Solomon Code over our construction of DNA codes provides a code rate of $0.46125$ as compared to the previous construction in \cite{a0} with code rate $0.145$.  Although the asymptotic code rate DNA code from Hamming codes is a little less than from the construction provided in \cite{a0}, our construction provides a better code rate for DNA codes of certain lengths. For the second-degree Hamming code, our construction provides a comparatively better code rate than previous codes in the literature. Furthermore, the map $\phi$ considered in the letter naturally leads to increased distances over the DNA alphabet with some codewords; for example, $0010060006$ and $0000000000$ are two $(10,8,3)$ Reed Solomon codewords over $\Z_{11}$ with $d_H(0000000000,0010060006)= 3$ but $d_H(f(\phi(0000000000)),f(\phi(0010060006))=5$. \\
{\bf Remark:} The DNA sequences constructed by the method provided in the letter have the unique characteristic of preserving the property of homopolymers of run length at most four and avoiding secondary structure of stem length more than two with concatenation operation. This operation enables us to construct families of larger lengths using concatenation operation, preserving the stated properties. This preservation of properties under the concatenation operation applies to the atomic families only and not to the families obtained by including code and its complement, such as the code in the 9th row of the table.

\section{Conclusions}
The methods discussed in the letter provide secondary structure free DNA codes with homopolymer run length less than four.  These codes have the code rates $0.5765 (\frac{log_4 11} {3})$ times the code rate of the corresponding  code over $\Z_{11}$. 
 For some specific lengths, they provide better code rates and satisfy more constraints than the existing codes in the literature. Although our method yields better performing codes with specific lengths, it is an interesting open question to determine an appropriate algebraic structure that leads to optimal codes.
\begin{center}
\label {tab}
\begin{table}
\begin{tabular}{|c|c|c|c|c|}
\hline
 $(n,M,d_H)$  & Code Rate& $P_1$ & $P_2$  & $P_3$ \\ 
DNA Codes &$log_4M/n$ & & &  \\
 \hline 
 \hline
 Example 9 \cite{a13} & $0.14937$  & No &$>4$ &No \\
 \hline 
 Example 2 \cite{a9} & $0.25850$  &Yes &$>4$ &No\\
 \hline
 Example 4 \cite{a9}& $0.40105$ &Yes & $>4$ &No\\
 \hline
 $(8,256,4)$ code& $0.50000$  & No&No &Yes\\
 Table III \cite{a4}& & & & \\
 \hline
 $(8,244,4)$ code& $0.48796$  &No &No &Yes\\
 Table III \cite{a4}& & & & \\
 \hline
 $f(\phi(H_7))$&$0.58027$  &Yes &$<4$ &Yes\\
 \cite{a0}& & & & \\
 \hline 
 $f(\phi(H_5))$& $0.57639$  &Yes &$<4$ &Yes\\
 $over \Z_{11}$& & & &\\
 \hline 
 $C \cup C^c$& $0.42865$  &Yes &$<4$ &Yes\\
 $C=f(\phi(H_2))$ & & & & \\
 \cite{a0}& & & & \\
 \hline
 $C \cup C^c$& $0.494365$  &Yes &$<4$ &Yes\\
 $C=f(\phi(H_2))$& & & & \\
 $over \Z_{11}$& & & & \\
 \hline
 $f(\phi(H_2))$&$0.42865$  &Yes& $<4$&Yes\\
 \cite{a0}& & & &\\
 \hline 
 $f(\phi(H_2))$&$0.48047$ &Yes& $<4$&Yes\\
 $over \Z_{11}$& & & &\\
 \hline  
 $C \cup C^c$&$0.35274$  & Yes&$<4$ &Yes \\ 
 $C=f(\phi (C_2))$& & & & \\ 
 \cite{a0}& & & & \\
 \hline 
 $f(\phi (C_2))$&$0.29024$  &Yes &$<4$ &Yes\\ 
 \cite{a0}& & & &\\
 \hline
 $f(\phi (C_2))$&$0.28828$  &Yes &$<4$ &Yes\\
 $over \Z_{11}$& & & &\\
 \hline
 $f(\phi((10,8,3)$&$0.145120$  & Yes&$<4$ &Yes\\
 $RS-code))$& & & &\\
 \cite{a0}& & & &\\
 \hline
 $f(\phi((10,8,3)$ &$0.461258$  &Yes & $<4$&Yes\\
 $RS-code))$& & & &\\
 \hline
\end{tabular}
\\
 \caption{ Comparisons of new codes with the codes in literature the properties: $P_1$ is the secondary structure avoiding property, $P_2$ is Homopolymer run length, $P_3$ is reverse complement constraint.}

\end{table}
\end{center}

\section*{Appendices}
\textbf{Proposition 1.} \textit{For any DNA sequence in $S^n$, $E_{1,3n} \geq -4n$}.\\ 
Proof of the above Proposition has been provided in the appendix.\\
\textit{Proof.} According to NJ algorithm, the minimum free energy for a subsequence $x_ix_{i+1}\cdots x_j$ of a DNA sequence $x_1x_2 \cdots x_n$ is given as
$$E_{i,j} = min\{E_{i+1,j-1}+\mu (x_i,x_j), E_{i,k-1}+E_{k,j} : k = i+1,\cdots,j\}$$ with initial conditions $E_{l,l} = E_{l-1,l} = 0$ for $l=1,2,\cdots,n$ \cite{a9}.\\
The commonly used value of $\mu(x_i,x_j)$ are
\[\mu (x_i,x_j) = 
\left\{
\begin{array}{ll}
      -5 & if (x_i,x_j)\in \{(C,G),(G,C)\}, \\
      -4 & if (x_i,x_j)\in \{(T,A),(A,T)\}, \\
      -1 & if (x_i,x_j)\in \{(T,G),(G,T)\}, \\
       \ \ 0 & otherwise \\
\end{array} 
\right. \]
For sequences of length $3n$ over $S^n$, the minimum free energy is
$$E_{1,3n} = min\{E_{2,3n-1}+\mu (x_1,x_3n), E_{1,k-1}+E_{k,3n} : k = 2,\cdots,j\}.$$
As the above equation is a recurrence relation depended on the interaction energy $\mu(x_i,x_j)$, for $1 \leq i \leq j \leq 3n$ and initial values of free energies are $0$. Free energy of any sequence is a function of interaction energy.
As the base alphabet set $S$ does not have any element with nucleotide $G$, the interaction energy can only be released by the interaction of nucleotide pairs $(A, T)$  and $(T, A)$. 
So,for any sequence $X$ of length $3n$, $$E_{1,3n}(X)= -4 \times \text{number of (A,T) pairs present in the sequence } X $$.
For n=1,
$E_{1,3}(CCC) = 0$, 
$E_{1,3}(CCA) = 0$, 
$E_{1,3}(CAC) = 0$, 
$E_{1,3}(CAA) = 0$, 
$E_{1,3}(ACC) = 0$, 
$E_{1,3}(ACA) = 0$, 
$E_{1,3}(AAC) = 0$, 
$E_{1,3}(AAA) = 0$, 
$E_{1,3}(TCC) = 0$, 
$E_{1,3}(CTC) = 0$, 
$E_{1,3}(TCA) = -4$.\\
So, for any DNA  sequence over $S$, $E_{1,3} \geq -4$.\\
For n=2, The sequence $TCATCA$ has only two pairs of (T, A) and $E_{1,6}(TCA TCA)=-8$. So, for any DNA  sequence over $S^2$, $E_{1,6} \geq -8$.\\
Let it be true for a positive integer k. Then for any DNA sequences over $S^k$, $E_{1,3k} \geq -4k$. On sequences over $S^k$, only the sequence $TCA \cdots TCA$ has $k$ pairs of (T, A) on it and \\$E_{1,3k} \underbrace{(TCA \cdots TCA)}_{\text{k times}} = -4k$.\\
For n=k+1,
The sequence $TCA \cdots TCA$ has the highest number of (T, A) pairs and  \\$E_{1,3(k+1)} \underbrace{(TCA \cdots TCA)}_{\text{k+1 times}} = -4(k+1)$. Hence for any DNA sequences over $S^k$, $E_{1,3(k+1)} \geq -4(k+1)$.\\
So the proposition is true for $n=k+1$, and hence by the method of mathematical induction, the proposition is true for any positive integer $n$. \hfill$\square$\\

%




\end{document}